\documentclass[prb,showpaces,preprintnumbers,amsmath,amssymb,twocolumn,superscriptaddress,longbibliography]{revtex4-1}
\usepackage{graphicx}
\usepackage{dcolumn}
\usepackage{bm}
\usepackage{amssymb}
\usepackage{amsmath}
\usepackage{epstopdf}
\usepackage{booktabs}
\usepackage{threeparttable}
\usepackage[pdfstartview=FitH, CJKbookmarks=true, bookmarksnumbered=true, bookmarksopen=true, colorlinks, pdfborder=001, linkcolor=blue, anchorcolor=blue, citecolor=blue]{hyperref}
\begin{document}
\title{Nodal multigap superconductivity in KCa$_2$Fe$_4$As$_4$F$_2$}
\author{M. Smidman}
\email{msmidman@zju.edu.cn}
\affiliation{Center for Correlated Matter and Department of Physics, Zhejiang University, Hangzhou 310058, China}
\author{F. K. K. Kirschner}
\affiliation{Department of Physics, University of Oxford, Clarendon Laboratory, Parks Road, Oxford OX1 3PU, United Kingdom}
\author{D. T. Adroja}
\email{devashibhai.adroja@stfc.ac.uk}
\affiliation{ISIS Facility, Rutherford Appleton Laboratory, Chilton, Didcot Oxon, OX11 0QX, United Kingdom} 
\affiliation{Highly Correlated Matter Research Group, Physics Department, University of Johannesburg, PO Box 524, Auckland Park 2006, South Africa}
\author{A. D. Hillier}
\affiliation{ISIS Facility, Rutherford Appleton Laboratory, Chilton, Didcot Oxon, OX11 0QX, United Kingdom} 
\author{F. Lang}
\affiliation{Department of Physics, University of Oxford, Clarendon Laboratory, Parks Road, Oxford OX1 3PU, United Kingdom}
\author{Z. C. Wang}
\affiliation{Department of Physics, Zhejiang University, Hangzhou 310058, China}
\author{G. H. Cao}
\affiliation{Department of Physics, Zhejiang University, Hangzhou 310058, China}
\author{S. J. Blundell}
\affiliation{Department of Physics, University of Oxford, Clarendon Laboratory, Parks Road, Oxford OX1 3PU, United Kingdom}
\date{\today}

\begin{abstract}
We find evidence that the newly discovered Fe-based superconductor  KCa$_2$Fe$_4$As$_4$F$_2$ ($T_c~=~33.36(7)$~K) displays multigap  superconductivity with line nodes. Transverse field muon spin rotation ($\mu$SR) measurements show that the temperature dependence of the superfluid density does not have the expected behavior of a fully-gapped superconductor, due to the  lack of saturation at low temperatures. Moreover, the data cannot be well fitted using either single band models or a multiband $s$-wave model, yet are well described by two-gap models with line nodes on either one or both of the gaps. Meanwhile the zero-field $\mu$SR results indicate a lack of time reversal symmetry breaking in the superconducting state, but suggest the presence of magnetic fluctuations. These results demonstrate a different route for realizing nodal superconductivity in iron-based superconductors. Here the gap structure is drastically altered  upon replacing one of the spacer layers, indicating the need to understand how the  pairing  state is tuned by changes of the asymmetry between the pnictogens located either side of the Fe planes.
\end{abstract}

\pacs{}
\maketitle

Following  the discovery of the second family of high temperature superconductors, the iron pnictides \cite{IronRep1,FeAsRep}, there has been considerable effort to understand the underlying mechanism for the formation of Cooper pairs. Identifying the pairing symmetry is one of the most important means of  determining the mechanism, for which it is vital to characterize the superconducting gap structure \cite{RoPP2011}. Many iron-arsenide-based superconductors have been proposed to have $s_\pm$ pairing symmetry where the superconductivity is mediated by spin fluctuations \cite{Mazin2008,Kuroki2008}. Here the superconducting gap remains fully open across the whole Fermi surface, but there is a change of sign of the gap between the hole pockets at the Brillouin zone center and the electron pockets at the zone edge. This scenario is well supported by measurements of the gap symmetry of many compounds, where evidence for two-gap nodeless superconductivity is found \cite{ding2008observation,Khasanov2009,Shermadini2010,Kim2011,Nakayama2011,Reid2016,1144Gap1,1144Gap2}. 

However, the universality of this picture was called into question by the observation of nodal superconductivity in a number of FeAs-based superconductors. These include the 1111 oxypnictides \cite{Mukuda2008,Grafe2008,Nakai2008}, BaFe$_2$(As$_{1-x}$P$_x$) \cite{Hashimoto2010,Nakai2010,YamashitaAsPNode,Qiu2012},  and KFe$_2$As$_2$ \cite{2009K122Nodes,2010K122Nodes2,2010K122Nodes,okazaki2012octet,K122ThermalCond}. The latter material corresponds to the heavily hole doped region of the phase diagram with a relatively low superconducting transition temperature of $T_c\approx 3$~K, as compared to the optimally doped Ba$_{0.6}$K$_{0.4}$Fe$_2$As$_2$ which has a higher value of $T_c\approx 37$~K and two nodeless gaps \cite{ding2008observation}. This crossover from nodeless to nodal superconductivity was suggested to correspond to a change from $s_{\pm}$ to $d$-wave pairing symmetry \cite{122dtheor2,122dtheor,K122ThermalCond}. Between these states the system would be expected to pass through a time reversal symmetry breaking $s+id$ state \cite{sidtheory}. Mixed evidence for such a phase has been found from  muon-spin relaxation ($\mu$SR), where time reversal symmetry breaking was not found in initial measurements  \cite{122TRS}, but evidence was subsequently observed from measurements of ion-irradiated samples \cite{122TRS2}. Meanwhile ARPES measurements indicate that KFe$_2$As$_2$ displays nodal superconductivity but with an $s$-wave pairing state overall \cite{okazaki2012octet}. 

The `122'  structure of (Ba,K)Fe$_2$As$_2$ is body centered tetragonal, where the Fe$_2$As$_2$ layers which are ubiquitous in the iron arsenide superconductors, are situated between layers of the alkaline/alkaline earth atoms. However, FeAs-based materials consisting of different layered arrangements can also be synthesized such as Ca$A$Fe$_4$As$_4$ ($A$~=~K, Rb, Cs) \cite{1144Rep}. In this case, the structure now has alternating sheets of Ca and $A$ atoms along the $c$-axis, which breaks the symmetry in the  Fe$_2$As$_2$  layers and the As atoms above and below the Fe-plane are no longer crystallographically equivalent. Meanwhile  stoichiometric CaKFe$_4$As$_4$ has a high $T_c$ of around 35~K, and is intrinsically near optimal hole doping \cite{1144Rep,1144Prop}. Measurements of the superconducting gap structure and inelastic neutron scattering  give clear evidence for a nodeless   $s_\pm$ pairing state, in line with the optimally doped `122' materials \cite{CaKFe4As4ARPES,1144Gap1,1144Gap2,1144INS,1144NMR}.

\begin{figure}[tb]
\begin{center}
 \includegraphics[width=0.99\columnwidth]{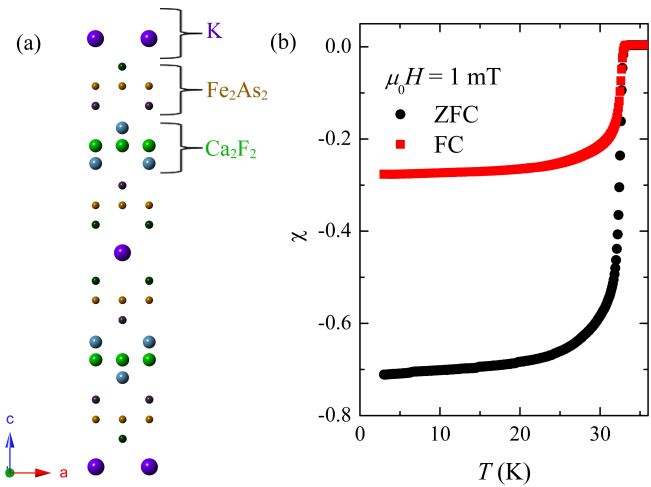}
\end{center}
\caption{ (a) Crystal structure of KCa$_2$Fe$_4$As$_4$F$_2$, where the K (purple), Fe (brown), As (dark green), Ca (blue) and F (light green) atoms are displayed. The structure consists of Fe$_2$As$_2$ layers which lie between K atoms on one side and Ca$_2$F$_2$ on the other. (b) Temperature dependence of the magnetic susceptibility of KCa$_2$Fe$_4$As$_4$F$_2$ for both zero-field cooled (ZFC) and field-cooled (FC) measurements. }
\label{Fig1}
\end{figure}

Another recently discovered variant showing high temperature superconductivity are  $A$Ca$_2$Fe$_4$As$_4$F$_2$ ($A$=K, Rb, Cs) \cite{12442Rep,1244rep2}. The crystal structure is displayed in Fig.~\ref{Fig1}(a), where the Fe$_2$As$_2$  layers are now surrounded by $A$ atoms on one side and Ca$_2$F$_2$ on the other, again leading to two distinct As sites above and below the Fe-plane. These materials also show large $T_c$ values of 28-33~K, depending on the element $A$, and are situated near to optimal doping. However the nature of the superconducting gap and therefore the pairing states of these new variants have not been characterized. In this Letter we report zero and transverse field $\mu$SR measurements of KCa$_2$Fe$_4$As$_4$F$_2$. The superfluid density derived from the depolarization rate of the transverse field measurements shows a lack of saturation at low temperatures, and the analysis provides clear evidence for nodal multigap superconductivity. Meanwhile, the  relaxation  observed in zero-field $\mu$SR shows a weak temperature dependence but no evidence for time reversal symmetry breaking upon entering the superconducting state. These results indicate a different route to nodal superconductivity in high temperature FeAs-based superconductors.

Polycrystalline samples of  KCa$_2$Fe$_4$As$_4$F$_2$ were synthesized using the solid state reaction method  described in Ref.~\onlinecite{12442Rep}. As can be seen from the temperature dependence of the magnetic susceptibility shown in Fig.~\ref{Fig1}(b), the samples show a sharp superconducting transition at around $T_c\approx33$~K. Muon-spin relaxation/rotation ($\mu$SR) measurements were performed on the MuSR spectrometer at the ISIS facility \cite{MuonREF}. Spin-polarized positive muons are implanted into the sample, which decay with a half life of 2.2~$\mu$s, emitting a positron. Since the positrons are preferentially emitted along the direction of the muon spin, by detecting the asymmetry of the emitted positrons, information can be obtained about the local magnetic field distribution at the muon stopping site. Zero-field $\mu$SR measurements were performed with the detectors in the longitudinal configuration, where the stray magnetic fields are cancelled to within 1~$\mu$T using an active compensation system. The transverse field measurements were performed with detectors in a transverse arrangement, with a field of 40~mT applied perpendicular to the initial muon polarization direction. The KCa$_2$Fe$_4$As$_4$F$_2$  sample was powdered and mounted on a silver plate (99.999$\%$), since the signal from muons stopping in  silver depolarizes at a negligible rate. All of the data were analyzed using WiMDA \cite{PRATT2000710}. 

\begin{figure}[t]
\begin{center}
 \includegraphics[width=0.9\columnwidth]{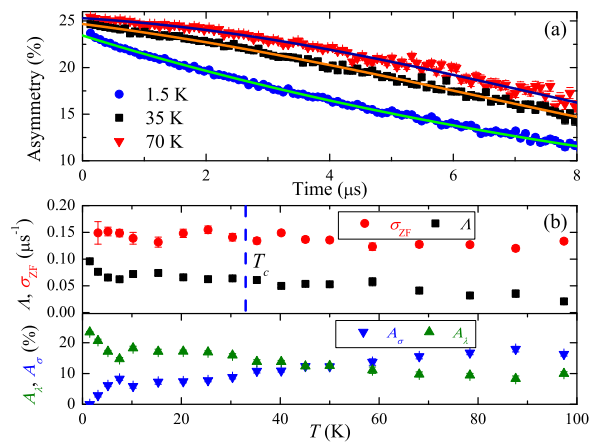}
\end{center}
\caption{(a) Zero-field $\mu$SR spectra at three temperatures, where the solid lines show fits to the data described in the text. (b) Temperature dependence of the Lorentzian ($\Lambda$) and Gaussian  ($\sigma_{{\rm ZF}}$) relaxation rates, along with the corresponding amplitudes of each component,  from fits to the zero-field $\mu$SR. }
\label{Fig2}
\end{figure}

Zero-field $\mu$SR measurements were performed from 1.5~K to 100~K, and the observed  asymmetries are displayed in Fig.~\ref{Fig2}(a) for three temperatures. The data were fitted with the sum of a  Lorentzian and Gaussian relaxation function $A(t)=A_{\lambda}{\rm exp}(-\Lambda t)+A_{\sigma}{\rm exp}(-\sigma^2_{{\rm ZF}}t^2/2) + A_{\rm bg}$, where the background term $A_{bg}$ was fixed from fitting at 100~K. It should also be noted that at 1.5~K, there is evidence for a small additional fast relaxation at short times. The fitted parameters are displayed in Fig.~\ref{Fig2}(b), where it can be seen  that with decreasing temperature, there is a gradual increase of $\Lambda$, whereas there is little change of $\sigma_{{\rm ZF}}$. These results suggest the presence of weak magnetic fluctuations, but  neither quantity shows a  detectable anomaly upon passing through $T_c$, indicating an absence of time reversal symmetry breaking. However, since $\Lambda(T)$ is not temperature independent, a small time reversal symmetry breaking signal as observed in some superconductors cannot be completely excluded \cite{Re6ZrTRS,K2Cr3As3MuSR}.

\begin{figure}[tb]
\begin{center}
 \includegraphics[width=0.99\columnwidth]{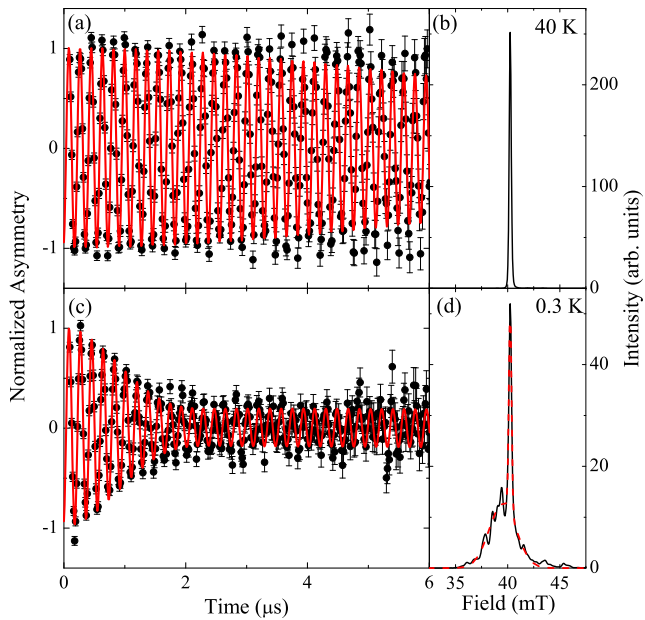}
\end{center}
\caption{Muon spin rotation ($\mu$SR) measurements of  KCa$_2$Fe$_4$As$_4$F$_2$ in a transverse field of 40~mT at (a) 40~K, where (b) displays the maximum entropy spectrum, and (c) at 0.3~K, with the corresponding maximum entropy plot shown in (d). The solid red lines in (a) and (c) show the fits described in the text, while the dashed line in (d) displays a fit with two Gaussian functions. }
\label{Fig3}
\end{figure}

Figures~\ref{Fig3}(a) and (c) display $\mu$SR spectra measured in a transverse field of 40~mT, performed at 40~K and 0.3~K respectively, above and below $T_c$. At 40~K, the muons precess at a single frequency, as shown by the maximum entropy spectrum in Fig.~\ref{Fig3}(b), with a slow depolarization arising from nuclear moments which are quasistatic on the timescale of the muon lifetime. Meanwhile  in the superconducting state  at 0.3~K, it can be seen that there is a significant increase of the depolarization rate [Fig.~\ref{Fig3}(c)]. The maximum entropy spectrum in Fig.~\ref{Fig3}(d) shows that in addition to a sharp narrow peak in the field distribution  centered around the applied field, there is also a broader component, the bulk of which corresponds to fields smaller than the applied transverse field. Such a distribution is a clear signature of a type-II superconductor in the mixed state \cite{BRANDT2009695}. 

\begin{figure}[tb]
\begin{center}
 \includegraphics[width=0.9\columnwidth]{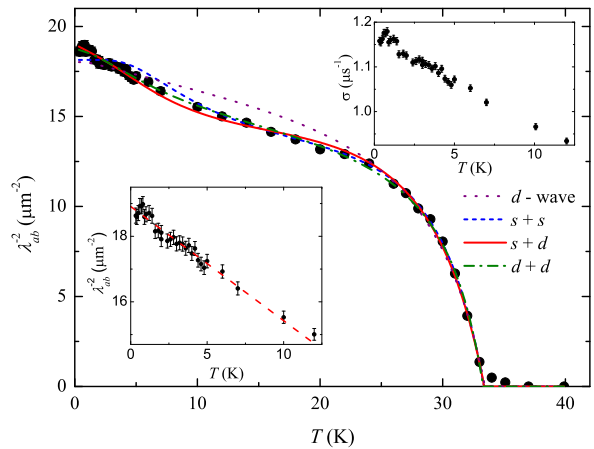}
\end{center}
\caption{Temperature dependence of the square of the inverse in-plane penetration depth $\lambda_{ab}^{-2}(T)$, which is proportional to the superfluid density. Fits to different models of the superconducting gap are displayed, where the single gap $d$-wave  and  fully gapped $s+s$ models cannot account for the data, while good fits are obtained for the two-gap nodal $s+d$ and $d+d$ models. The insets shows the low temperature behavior of both $\lambda_{ab}^{-2}(T)$ and the total Gaussian relaxation rate $\sigma$, where the $\lambda_{ab}^{-2}(T)$ data up to 10~K is consistent with a linear temperature dependence.}
\label{Fig4}
\end{figure}

As shown by the dashed line in Fig.~\ref{Fig3}(d), the field distribution in the superconducting state can be well accounted for by fitting with two Gaussian functions. Since the narrow component at the applied field corresponds to muons stopping in the silver sample holder, this demonstrates that the field distribution within the vortex lattice is well characterized by a single Gaussian centered at a field below 40~mT. Consequently, the time spectra were fitted using 

\begin{equation}
A(t)=A_0{\rm e}^{-\sigma^2t^2/2}{\rm cos}(\gamma_{\mu}B_0t + \phi)+ A_1{\rm cos}(\gamma_{\mu}B_1t + \phi), 
\label{TFFit}
\end{equation}

\noindent where $\gamma_{\mu}/2\pi=135.5$~MHz/T is the muon gyromagnetic ratio, $\sigma$ is the Gaussian relaxation rate, $\phi$ is related to the detector geometry, $A_0$ and $A_1$ are the amplitudes of the components from the sample and silver respectively, while $B_0$ and $B_1$ are the corresponding internal magnetic fields. The total amplitudes for each group of detectors were fixed, as well as the amplitude ratio $A_0/A_1=4.12$,  allowing for the temperature dependence of the muon relaxation rate $\sigma(T)$ to be obtained. We note that adding an additional Gaussian term to Eq.~\ref{TFFit} leads to overfitting of the data, and therefore this was not used. In addition, unlike the zero-field measurements, the transverse field spectra are well described at all temperatures by a purely Gaussian relaxation, with no Lorentzian component. The superconducting contribution to the depolarization rate was calculated by subtracting the nuclear contribution via $\sigma_{sc}=\sqrt{\sigma^2-\sigma_{{\rm nuc}}^2}$, where the component from the nuclear spins  $\sigma_{{\rm nuc}}=0.092(1)\mu$s$^{-1}$ was determined from the values above $T_c$.  For applied fields much less than the upper critical field,  $\sigma_{sc}$ can be related to the effective penetration depth $\lambda_{{\rm eff}}$ using $\sigma_{sc}/\gamma_{\mu} = 0.0609\Phi_0/\lambda_{{\rm eff}}^2$, where $\Phi_0$ is the magnetic flux quantum \cite{Brandt2003}. Since the material is an anisotropic layered compound where the out of plane penetration depth ($\lambda_c$) greatly exceeds the in-plane value ($\lambda_{ab}$), for a polycrystalline sample $\lambda_{{\rm eff}}$ is dominated by $\lambda_{ab}$, where $\lambda_{{\rm eff}}=3^{\frac{1}{4}}\lambda_{ab}$ \cite{FESENKO1991551}.

Figure~\ref{Fig4} displays the temperature dependence of $\lambda_{ab}^{-2}(T)$, which is proportional to the superfluid density and therefore provides information about the superconducting gap structure. It can be seen that with decreasing temperature there is no evidence for the saturation  of $\lambda_{ab}^{-2}(T)$ down to the lowest measured temperatures, indicating the presence of nodes in the superconducting gap. For a fully gapped superconductor at sufficiently low temperatures, thermal excitations are unable to depopulate the superconducting condensate, leading to constant $\lambda_{ab}^{-2}(T)$. Meanwhile if there are nodes in the  gap, there are always  low energy  excitations comparable to the thermal energy and hence $\lambda_{ab}^{-2}(T)$ will continue to increase upon lowering the temperature. Furthermore, if the gap contains lines of nodes, at low temperatures the temperature dependence of the  penetration depth is linear. Therefore for sufficiently low temperatures $\lambda_{ab}^{-2}(T)\approx\lambda_{ab}^{-2}(0)-aT$, where $a$ is a constant, and it can be seen in the inset that the data are compatible with a linear increase from the lowest measured temperature up to around 10~K (corresponding to a total change in $\sigma(T)$ of $\approx0.2\mu s^{-1}$).

The normalized superfluid density $\tilde{n}(T)=[\lambda_{ab}(T)/\lambda_{ab}(0)]^{-2}$ was modelled using \cite{SuperFRef}

\begin{equation}
\tilde{n}(T) = 1 + \frac{1}{\pi} \int_{0}^{2\pi} \int_{\Delta(T, \varphi)}^{\infty}\frac{\partial f}{\partial E}\frac{E{\rm d}E{\rm d}\varphi}{\sqrt{E^2-\Delta^2(T, \varphi)}},
\label{RhoS}
\end{equation}

\noindent where $f~=~\left[1+\exp\left(-E/k_{\mathrm{B}}T\right)\right]^{-1}$  is the Fermi-Dirac function. The gap function $\Delta(T, \varphi)~=~\Delta(T)g(\varphi)$ has a temperature dependence given by  $\Delta(T)=\Delta(0)$tanh$[(1.82){(1.018(\it {T}_c/T-1))}^{0.51}]$ \cite{CarringtonMgB2}, where $\Delta(0)$ is the zero temperature magnitude. The angular dependence $g(\varphi)$ is given by $g(\varphi)=1$ or ${\rm cos}(2\varphi)$ for an $s$-wave ($\Delta^s$) or $d$-wave ($\Delta^d$) gap respectively ($\varphi$~=~azimuthal angle).

The data were fitted with various models for the superfluid density, with both one and two gaps. Neither a single fully gapped $s$-wave model (not displayed), nor a single $d$-wave gap with line nodes can fit the data well. In particular, it can be seen in Fig.~\ref{Fig4} that upon reducing the temperature, there is an inflection point at around 10~K, below which there is an upturn in  $\lambda_{ab}^{-2}(T)$. Such behavior is difficult to account for with single gap models, but suggests the presence of multiple gaps. Various two-gap models were fitted to the data by adding the weighted sum of two components, where  $\tilde{n}(T)=x\tilde{n}_1^{s,d}(T)+(1-x)\tilde{n}_2^{s,d}(T)$, where $\tilde{n}_i^{s}(\tilde{n}_i^{d})$ is the superfluid density corresponding to a gap $\Delta_i^s$ ($\Delta_i^d$), with a weight for the $i=1$ component $0\leq x \leq 1$. A model with two isotropic gaps ($s+s$) can also not well describe the data due to the lack of a low temperature plateau as discussed previously. Meanwhile both an $s+d$ model, with one fully open gap and one line nodal gap, as well as a $d+d$ model with two nodal $d$-wave gaps, provide good fits to the superfluid density. The fitted   parameters are  $\Delta^s_1(0)=10.12(7)$~meV, $\Delta_2^d(0)=1.84(2)$~meV, $x=0.70(1)$ and $T_c=33.36(7)$~K for the $s+d$ model, while for $d+d$ they are $\Delta_1^d(0)=1.71(3)$~meV, $\Delta_2^d(0)=14.6(3)$~meV, $x=0.15(1)$ and $T_c=33.28(3)$~K. These fits yield respective zero temperature penetration depth values of  $\lambda_{ab}(0)=229.5(5)$ and 229.8(5)~nm. Therefore our analysis of the superfluid density from transverse field $\mu$SR indicates the presence of multigap nodal superconductivity.

The observation of multiband nodal superconductivity in  KCa$_2$Fe$_4$As$_4$F$_2$ is markedly different to the similar iron pnictide superconductor CaKFe$_4$As$_4$, where clear evidence is found for multigap nodeless superconductivity with an $s_{\pm}$ pairing state \cite{CaKFe4As4ARPES,1144Gap1,1144Gap2,1144INS,1144NMR}. This difference is all the more puzzling since both materials are stoichiometric compounds and therefore should be similarly near optimal hole doping \cite{12442Rep,1144Prop}. In addition, similar to many other FeAs-based superconductors but unlike the nodal material KFe$_2$As$_2$ \cite{okazaki2012octet}, electronic structure calculations for  KCa$_2$Fe$_4$As$_4$F$_2$ show the presence of hole pockets at the zone center and electron pockets at the edge, from which a nodeless $s_{\pm}$ state may be anticipated \cite{12442Calc}. Therefore the change in gap structure between CaKFe$_4$As$_4$ and  KCa$_2$Fe$_4$As$_4$F$_2$ is not likely to be analogous to the case of (Ba$_{1-x}$K$_x$)Fe$_2$As$_2$, where the shift from nodeless to nodal superconductivity emerges upon strong hole doping as $x$ is increased.  There is also evidence that pressure can lead to a change from nodeless to nodal superconductivity in Ba$_{0.65}$Rb$_{0.35}$Fe$_2$As$_2$ \cite{guguchia2015direct}, and it was proposed that gap nodes can emerge in several iron pnictide superconductors when the height of the pnictogen above the Fe-plane falls below 1.33~\AA~\cite{NodalPnHeight}. However, in KCa$_2$Fe$_4$As$_4$F$_2$ the As atoms are at heights of 1.40 and 1.44~\AA~\cite{12442Rep}, well above the proposed upper limit and even above the values found for nodeless CaKFe$_4$As$_4$ (1.40 and 1.35~\AA) \cite{1144Rep}.

For the  fits to the superfluid density of  KCa$_2$Fe$_4$As$_4$F$_2$ using the  $s+d$ and $d+d$ models, the respective magnitudes of the larger of the two gaps are $3.52(3)k_{\rm B}T_c$ and $5.08(1)k_{\rm B}T_c$, which we note are significantly greater than the  theoretical weak coupling values ($1.76k_{\rm B}T_c$ and $2.14k_{\rm B}T_c$ for $s$- and $d$-wave respectively) \cite{swaveWC,dwaveWC}. While this may suggest the presence of strongly coupled superconductivity, the fitted gap values for models with  highly anisotropic gaps are sensitive to the form of the gap function and the nature of the Fermi surface. For instance, if the gap nodes are only present on a relatively small region of the Fermi surface, as was proposed for BaFe$_2$(As$_{0.7}$P$_{0.3}$)$_2$ from ARPES measurements \cite{zhang2012nodal}, then the superfluid density would drop less rapidly with temperature than for a $d$-wave gap with more extended nodal regions, which may account for the larger extracted gap values. Such a small nodal region is also consistent with the high value of $T_c$ \cite{zhang2012nodal}, as compared to the considerable lower value in KFe$_2$As$_2$  ($T_c\approx 3$~K). However, while in both BaFe$_2$(As$_{1-x}$P$_{x}$)$_2$ and KCa$_2$Fe$_4$As$_4$F$_2$ the nodal superconductivity does not arise due to tuning the carrier concentration, in BaFe$_2$(As$_{1-x}$P$_{x}$)$_2$, increasing $x$ corresponds to doping into the Fe-As layers leading to a positive chemical pressure effect \cite{Jiang2009}. This is quite different to the case of  CaKFe$_4$As$_4$ and KCa$_2$Fe$_4$As$_4$F$_2$, where the nodal superconductivity is brought about by swapping one of the spacer layers. It is noted that  both these  materials are asymmetric with respect to the Fe atoms, so that the As above and below the Fe-planes are crystallographically inequivalent,  and the effect of this asymmetry on the electronic structure and pairing symmetry requires further exploration.  However, our results indicate a different route for tuning the gap structures of iron based superconductors. As such, in order to understand this change and to identify the pairing symmetry, it is vital to further characterize the superconducting properties of  KCa$_2$Fe$_4$As$_4$F$_2$. It is of particular importance to probe the exact nature of the gap anisotropy from single crystal studies and to look for the spin resonance via  inelastic neutron scattering measurements.

\begin{acknowledgments}
This work is supported by  the National Key R\&D Program of China (Grant No.~2017YFA0303100) and EPSRC grant EP/N023803. F.K.K.K. thanks Lincoln College, Oxford, for a doctoral studentship. D.T.A. would like to thank to the Royal Society of London for support from the UK-China Newton Mobility Grant.
\end{acknowledgments}

\end{document}